\documentclass[prc,letterpaper,onecolumn,showpacs,showkeys,lengthcheck,
               floatfix,nofootinbib,preprintnumbers,superscriptaddress]{revtex4-1} 

\usepackage{mathtools}
\usepackage{amssymb,amsfonts}
\usepackage{bm}
\usepackage{slashed}

\usepackage[svgnames]{xcolor}
\usepackage[english]{babel}
\usepackage{blindtext}
\usepackage{microtype}
\usepackage{tikz}
\usepackage{dcolumn}
\usepackage{multirow}
\usepackage[]{units}

\usepackage{graphicx}
\usepackage[caption=false]{subfig}

\usepackage[colorlinks=true]{hyperref}
\usepackage{ifpdf}

\def\a{\alpha}
\def\b{\beta}

\def\g{\gamma}

\def\ve{\varepsilon}

\def\l{\lambda}
\def\s{\sigma}

\def\D{\Delta}
\def\G{\Gamma}
\def\L{\Lambda}

\def\S{\Sigma}

\def\pl{\partial}
\def\hs{\hspace}

\def\ol{\overline}
\def\no{\nonumber}

\def\lf{\left}
\def\rg{\right}

\newcommand{\vect}[1]{\boldsymbol{#1}}

\newcommand{\sh}[1]{\slashed{#1}}

\font\bb=bbmss10 scaled 1200
\def\ident{\mbox{\bb 1}}

\def\be{\begin{equation}}
\def\ee{\end{equation}}

\begin{document}

\title{SU(3)-flavour breaking in octet baryon masses and axial couplings}

\author{Manuel~E.~Carrillo-Serrano}
\affiliation{CSSM and ARC Centre of Excellence for Particle Physics at the Tera-scale,\\
School of Chemistry and Physics,
University of Adelaide, Adelaide SA 5005, Australia
}

\author{Ian~C.~Clo\"et}
\affiliation{Physics Division, Argonne National Laboratory, Argonne, Illinois 60439, USA
}

\author{Anthony~W.~Thomas}
\affiliation{CSSM and ARC Centre of Excellence for Particle Physics at the Tera-scale,\\
School of Chemistry and Physics,
University of Adelaide, Adelaide SA 5005, Australia
}

\begin{abstract}
The lightest baryon octet is studied within a covariant and confining 
Nambu--Jona-Lasinio model. By solving the relativistic Faddeev equations 
including scalar and axialvector diquarks, we determine the masses and 
axial charges for $\Delta S = 0$ transitions. For the latter the 
degree of violation of SU(3) symmetry arising because of the strange 
spectator quark(s) is found to be up to 10\%.
\end{abstract}


\maketitle
\section{Introduction}
In the quest to fully understand Quantum Chromodynamics (QCD) it is not  
sufficient to study baryons whose valence quark content consists 
only of the light $u$ and $d$ quarks. A solid understanding of all 
members of the  baryon octet -- that is, the nucleon, $\Lambda$, $\Sigma$ 
and $\Xi$ multiplets -- remains a critical step. Early work on their 
structure centered on the  constituent quark 
model~\cite{LeYaouanc:1976ne,Isgur:1979be}  and the MIT bag 
model~\cite{Chodos:1974pn}, later  supplemented by chiral corrections 
associated with the cloud of virtual  pions and kaons that surround a  
baryon~\cite{Theberge:1980ye,Theberge:1981pu,Myhrer:1980jy,Myhrer:1982sp,
Tsushima:1988xv,Weigel:1993zd,Wagner:1998fi,Diakonov:1997sj,Cloet:2002eg}.  
Once their basic properties, such as masses, charge radii, magnetic moments  
and axial charges had been calculated,   attention naturally turned 
to more complex properties, such as their parton 
distribution functions~\cite{Boros:1999tb,Diakonov:1996sr}. 

The empirical evidence concerning the structure of the hyperons is 
naturally  far more limited than for nucleons. While we have fairly 
good data for their  masses, magnetic moments and axial 
charges~\cite{Beringer:1900zz}, little or nothing is known  concerning 
their electromagnetic or axial form factors as a function of  
momentum transfer. Finding ways to explore these properties experimentally  
would be very valuable. On the other hand,  over the last couple of 
decades lattice QCD has made  steady progress in the calculation of 
octet baryon  masses~\cite{Durr:2008zz,Young:2009zb,WalkerLoud:2008bp,Aoki:2008sm},
including determinations of their isospin mass splittings~\cite{Borsanyi:2014jba}, 
and certain  electroweak matrix elements~\cite{Lin:2007ap}.  
These studies have been complemented by  a judicious use of chiral 
effective field theory in order to  extrapolate to the physical quark 
masses. Thus, we now have quite accurate  determinations of the 
hyperon electric~\cite{Shanahan:2014cga}  and 
magnetic~\cite{Shanahan:2014uka} form factors up to  1.4 GeV$^2$, 
as well as low moments of their parton  
distribution functions~\cite{Cloet:2012db}.  
It has even been possible recently to shed  some light on the 
proton spin puzzle~\cite{Ashman:1987hv,Myhrer:2007cf} by calculating 
the spin fractions carried by quarks across the octet~\cite{Shanahan:2013apa}.

On general grounds one would prefer to have models of octet baryon properties  that are covariant as well as respecting the symmetries of QCD.  The former is especially important if one wants to investigate parton  distributions and form factors at high momentum transfer. The hope in building  more sophisticated models is that through comparison with empirical data  and lattice QCD studies one may develop a deeper understanding of how QCD  works in the non-perturbative regime, including issues such as the importance  and role of diquark correlations and chiral corrections~\cite{Cloet:2014rja},  as well as the transition from non-perturbative to  perturbative QCD~\cite{Cloet:2013gva}.   

In this work we investigate the masses and $\Delta S = 0$ axial charges 
of the octet baryons within the framework of the covariant model of 
Nambu and Jona-Lasinio (NJL)~\cite{Nambu:1961tp,Nambu:1961fr,Vogl:1991qt,Hatsuda:1994pi,Klevansky:1992qe}, 
where confinement is simulated by employing proper-time 
regularization~\cite{Ebert:1996vx,Hellstern:1997nv,Bentz:2001vc}. 
Octet baryons are described by a Poincar\'e covariant Faddeev equation,  
where scalar and axialvector diquarks correlations are assumed 
to play a dominant role. Flavour breaking effects, introduced by a dressed strange
quark that is approximately 50\% heavier that the dressed light quarks, will also be studied.

The structure of the paper is as follows.  Sect.~\ref{sec:NJL} provides a brief introduction to the NJL model,  including a discussion of the Bethe-Salpeter equation for mesons and diquarks.  Sect.~\ref{sec:baryons} introduces the Faddeev equation for octet baryons,  discussing the solution for the Poincar\'e covariant Faddeev amplitude  and octet masses. Finally, in Sect.~\ref{AxialCh}, the formalism is used  to determine the axial charges associated with strangeness conserving  beta decays. Sect.~\ref{sect:con} summarises our findings and presents  some concluding remarks.

\section{Nambu--Jona-Lasinio Model \label{sec:NJL}}

The NJL model was formulated as a theory of elementary fermions which  encapsulated dynamical chiral symmetry breaking in a transparent  manner~\cite{Nambu:1961tp,Nambu:1961fr}. With the advent of QCD, it was reformulated with quarks as the fundamental degrees of freedom, such that the symmetries of QCD are respected.\footnote{The $SU(3)$ colour gauge symmetry of QCD is a global symmetry of the NJL model.} In particular the NJL model exhibits dynamical chiral symmetry breaking, which, as implemented in this work, gives rise to approximately $95$\% of the nucleon mass. 
 
The complete three-flavour NJL Lagrangian in the $\bar{q}q$ interaction channel -- including only 4-fermion interactions -- has the form~\cite{Klevansky:1992qe}
\begin{align}
\mathcal{L} &= \bar{\psi}\lf(i\sh{\partial} - \hat{m}\rg)\psi \no \\
&+ \tfrac{1}{2}\,G_\pi\Big[\tfrac{2}{3}\lf(\bar{\psi}\psi\rg)^2 + \lf(\bar{\psi}\,\vect{\l}\,\psi\rg)^2 \no \\
&\hs*{35mm}
- \tfrac{2}{3}\lf(\bar{\psi}\,\g_5\,\psi\rg)^2 - \lf(\bar{\psi}\,\g_5\,\vect{\l}\,\psi\rg)^2\Big] \no \\
&- \tfrac{1}{2}\,G_\rho\lf[\lf(\bar{\psi}\,\g^\mu\,\vect{\l}\,\psi\rg)^2
+ \lf(\bar{\psi}\,\g^\mu\g_5\,\vect{\l}\,\psi\rg)^2\rg] \no \\
&
- \tfrac{1}{2}\,G_0\lf(\bar{\psi}\,\g^\mu\,\psi\rg)^2
- \tfrac{1}{2}\,G_5\lf(\bar{\psi}\,\g^\mu\g_5\,\psi\rg)^2,
\label{eq:lagrangian}
\end{align}
where $\vect{\l}$ represents the eight the Gell-Mann matrices and $\hat{m} = \text{diag}\lf[m_u,\,m_d,\,m_s\rg]$. The NJL model does not include gluons as explicit degrees of freedom, as such the pointlike quark--quark interaction renders the NJL model non-renormalizable. We regularize the NJL model using the proper-time scheme, which mantains Lorentz and gauge invariance, it also removes unphysical thresholds for the decay of colour singlet bound states in their coloured constituents, thereby simulating quark confinement~\cite{Ebert:1996vx,Hellstern:1997nv,Bentz:2001vc}.

\begin{figure}[tbp]
\centering\includegraphics[width=\columnwidth,clip=true,angle=0]{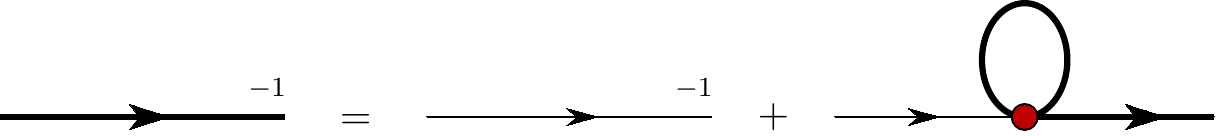}
\caption{(Colour online) The NJL gap equation in the Hartree-Fock approximation, where the thin line represents the elementary quark propagator, $S_{0q}^{-1}(k) = \sh{k} - m_q + i\ve$, and the shaded circle represents the 4-fermion interaction.}
\label{fig:gapequation}
\end{figure}

The dressed quark propagator in the NJL model is obtained from the gap equation illustrated in Fig.~\ref{fig:gapequation}. The solution for a quark of flavour $q = u,\,d,\,s$ has the form
\begin{align}
S_q(k)^{-1} = \sh{k} - M_q  + i\ve,
\end{align}
where, in the proper-time regularization scheme, the dressed quark masses each satisfy
\begin{align}
M_q = m_q + \frac{3}{\pi^{2}}\,M_q\,G_{\pi}\int_{1/\L_{UV}^{2}}^{1/\L_{IR}^{2}} d\tau\, \frac{e^{-\tau M_q^2}}{\tau^2}.
\label{gap}
\end{align}
In this three-flavour NJL model, defined by Eq.~\eqref{eq:lagrangian}, the gap equation does not introduce flavour mixing in the quark propagator, this is in contrast to the two-flavour case which in general has flavour mixing~\cite{Klevansky:1992qe}. 

The quark-quark interaction needed for the two-body interaction kernel in the Faddeev equation (to be described shortly) can be obtained from Eq.~\eqref{eq:lagrangian} using Fierz transformations. Keeping only scalar and axialvector diquark correlations, the NJL interaction Lagrangian in the $qq$ channel reads
\begin{align}
\mathcal{L}_{I}^{qq} &= G_s \Bigl[\bar{\psi}\,\g_5\, C\,\l_a\,\beta_A\, \bar{\psi}^T\Bigr]
                              \Bigl[\psi^T\,C^{-1}\g_5\,\l_a\,\beta_A\, \psi\Bigr] \no \\
&
+ G_a \Bigl[\bar{\psi}\,\g_\mu\,C\,\l_s\,\beta_A\, \bar{\psi}^T\Bigr]
                              \Bigl[\psi^T\,C^{-1}\g^{\mu}\,\l_s\, \beta_A\, \psi\Bigr],
\label{eq:qqlagrangian}
\end{align}
where $C = i\g_2\g_0$ is the charge conjugation matrix and the couplings 
$G_s$ and $G_a$ give the strength of the scalar and axialvector 
$qq$ interactions (with $G_s = -1/3G_{\rho} + 1/18G_0 + 1/18G_5$ and
$G_a = -1/6G_{\rho} - 1/18G_0 - 1/18G_5$). The flavour matrices are labelled by 
$\l_a=\l_2,\,\l_5,\,\l_7$ and $\l_s = \l_0,\,\l_1,\,\l_3,\,\l_4,\,\l_6,\,\l_8$,
where $\l_0 \equiv \sqrt{\frac{2}{3}}\,\ident$. Thus, there are three types 
of scalar and six types of axialvector diquarks. The color $\bar{3}$ 
matrices are given 
by $\beta_A = \sqrt{\tfrac{3}{2}}\,\lambda_A~(A=2,5,7)$~\cite{Ishii:1995bu,Ishii:1993np,Ishii:1993rt} 
and hence the interaction terms in Eq.~\eqref{eq:qqlagrangian} are 
totally antisymmetric, as demanded by the Pauli principle.

\begin{figure}[tbp]
\centering\includegraphics[width=\columnwidth,clip=true,angle=0]{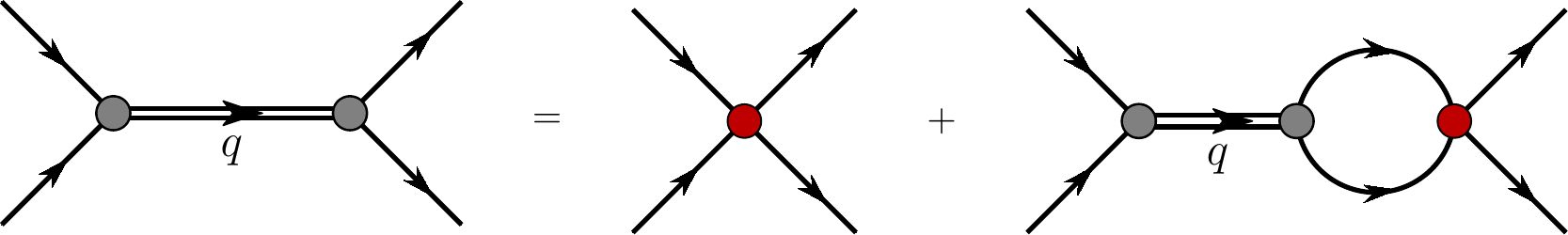}
\caption{(Colour online) Inhomogeneous Bethe-Salpeter equation for quark--quark (diquark) correlations.}
\label{fig:1}
\end{figure}

Quark-antiquark and quark-quark bound states are obtained by solving the appropriate Bethe-Salpeter equation, which is illustrated in Fig.~\ref{fig:1} for diquarks. The reduced $t$-matrices for scalar and axial-vector diquarks, with quark flavour content $q_1$ and $q_2$, take the form\footnote{Throughout this paper $[q_1q_2]$ will indicate a quantity associated with a scalar diquark and $\{q_1q_2\}$ will indicate an object associated with an axialvector diquark quantity.}
\begin{align}
\label{eq:tscalar}
\tau_{[q_1q_2]}(q) &= \frac{-4i\,G_s}{1 + 2\,G_s\,\Pi_{[q_1q_2]}(q^2)}, \\[0.7ex]
\label{eq:taxial}
\tau^{\mu\nu}_{\{q_1q_2\}}(q) 
&= \frac{-4i\,G_a}{1+2\,G_a\,\Pi_{\{q_1q_2\}}(q^2)} \no \\
&\hs{12mm} \times
\lf[g^{\mu\nu} + 2\,G_a\,\Pi_{\{q_1q_2\}}(q^2)\,\frac{q^\mu q^\nu}{q^2}\rg].
\end{align}
The bubble diagrams are given by
\begin{align}
\label{eq:bubble_PP}
&\Pi_{[q_1q_2]}\lf(q^2\rg) = \no \\
&\hs{17mm}6i \int \frac{d^4k}{(2\pi)^4}\ \mathrm{Tr}\lf[\g_5\,S_{q_1}(k)\,\g_5\,S_{q_2}(k+q)\rg], \\[0.5ex]
\label{eq:bubble_VV}
&\Pi_{\{q_1q_2\}}(q^2)\lf(g^{\mu\nu} - \frac{q^\mu q^\nu}{q^2}\rg) = \no \\
&\hs{17mm}6i \int \frac{d^4k}{(2\pi)^4}\ \mathrm{Tr}\lf[\g^\mu\,S_{q_1}(k)\,\g^\nu\,S_{q_2}(k+q)\rg],
\end{align}
where the flavour and colour traces have been taken, and the remaining trace is over Dirac indices only. The masses of the various diquarks are given by the poles in the corresponding $t$-matrix, e.g., the scalar diquarks masses are given by the pole condition
\begin{align}
1 + 2\,G_s\,\Pi_{[q_1q_2]}\bigl(q^2 = M_{[q_1q_2]}^2\bigr) = 0.
\end{align}

For the octet baryon calculations we approximate the full diquark $t$-matrix by a \textit{contact + pole} form, that is
\begin{align}
\label{eq:scalarpropagatorpoleform}
\tau_{[q_1q_2]}(q) &\to  4i\,G_s - \frac{i\,Z_{[q_1q_2]}}{q^2 - M_{[q_1q_2]}^2+i\,\ve}, \\
\label{eq:axialpropagatorpoleform}
\tau^{\mu\nu}_{\{q_1q_2\}}(q) &\to  \no \\
&\hs{-8mm}4i\,G_a - \frac{i\,Z_{\{q_1q_2\}}}{q^2 - M_{\{q_1q_2\}}^2+i\,\ve} \lf(g^{\mu\nu} - \frac{q^{\mu}q^{\nu}}{M_{\{q_1q_2\}}^2}\rg),
\end{align}
where the pole residues are given by
\begin{align}
\label{eq:Zs}
Z_{[q_1q_2]}^{-1}  &= -\frac{1}{2}\,\frac{\pl}{\pl q^2}\,\Pi_{[q_1q_2]}(q^2)\Big\rvert_{q^2 = M_{[q_1q_2]}^2}, \\[0.2ex]
Z_{\{q_1q_2\}}^{-1} &= -\frac{1}{2}\,\frac{\pl}{\pl q^2}\,\Pi_{\{q_1q_2\}}(q^2)\Big\rvert_{q^2 = M_{\{q_1q_2\}}^2}.
\end{align}

\section{Faddeev Equations for Octet Baryons \label{sec:baryons}}
\begin{figure}[tbp]
\centering\includegraphics[width=\columnwidth,clip=true,angle=0]{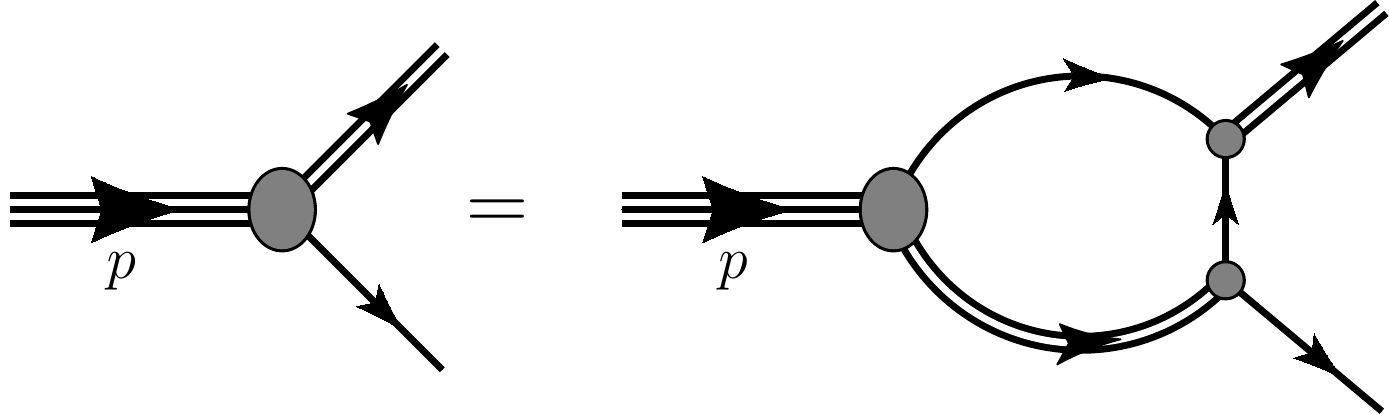}
\caption{Homogeneous Poincar\'e covariant Faddeev equation whose solution gives the mass and 
vertex function for each member of the baryon octet.}
\label{fig:2}
\end{figure}

Octet baryons are constructed as solutions to a Poincar\'e covariant Faddeev equation, which 
is illustrated in Fig.~\ref{fig:2}, where the quark--diquark approximation used here has 
been made explicit~\cite{Afnan:1977pi}. A tractable solution to the Faddeev equation is 
obtained by employing the static approximation~\cite{Buck:1992wz} to the quark exchange kernel, 
where the exchanged quark propagator becomes $S_q(k) \to -\frac{1}{M_q}$. This approximation
has been shown to yield excellent results for nucleon form factors~\cite{Cloet:2014rja} and 
quark distributions~\cite{Cloet:2005pp,Cloet:2005rt,Cloet:2007em,Bentz:2007zs}. The Faddeev equation for each octet baryon 
then takes the form
\begin{align}
\G_{B}(p,s) = Z_B\,\Pi_B(p)\,\G_{B}(p,s),
\label{eq:faddeev}
\end{align}
where $B$ labels an octet baryon, $Z_B$ the corresponding quark exchange kernel and $\Pi_B(p)$
is a diagonal matrix containing the various combinations of quark and diquark propagator.
Eq.~\eqref{eq:faddeev} must be supplemented by a normalization condition, such that
the normalized Faddeev vertex reads
\begin{align}
\G_{B}(p,s) = \sqrt{-\mathcal{Z}_B}\ \G_{0B}(p,s),
\end{align}
where $\G_{0B}(p,s)$ is the unnormalized vertex and the normalization condition that determines
$\mathcal{Z}_B$ will be discussed shortly.

For equal light quark masses the nucleon, $\Sigma$ and $\Xi$
Faddeev vertex functions contain one scalar diquark and two types 
of axial-vector 
diquark\footnote{For the nucleon, in the $M_u = M_d$ limit, 
the singly and doubly represented
axialvector diquarks are mass degenerate and could therefore be 
treated as a single type of diquark.
However, for the nucleon we will keep the description more general 
so that the analogy with the other members of the octet
is straightforward.}, with a Dirac structure of the form
\begin{align}
\G_b(p,s) &= \begin{bmatrix}
\G_{q_1[q_1q_2]}(p,s) \\[1.1ex]
\G_{q_1\{q_1q_2\}}^\mu(p,s) \\[1.1ex]
\G_{q_2\{q_1q_1\}}^\mu(p,s)
\end{bmatrix}, \\
&= \sqrt{-\mathcal{Z}_b}\,\begin{bmatrix} 
\a_1 \\[0.8ex]
\a_2\,\frac{p^\mu}{M_b}\,\g_5 + \a_3\,\g^\mu\g_5 \\[0.8ex]
\a_4\,\frac{p^\mu}{M_b}\,\g_5 + \a_5\,\g^\mu\g_5 
\end{bmatrix}u_b(p,s),  
\label{eq:vertex}
\end{align}
where $b = [\text{nucleon},\,\Sigma,\,\Xi]$ and $\mathcal{Z}_b$ is the 
vertex normalization. The Faddeev vertex function for the $\L$ baryon, with
equal $u$ and $d$ quark masses, contains two types of scalar diquark and 
an axial-vector diquark and therefore reads
\begin{align}
\G_\L(p,s) &= \sqrt{-\mathcal{Z}_\L}\,\begin{bmatrix} 
\a_1 \\[0.6ex]
\a_2 \\[0.6ex]
\a_3\,\frac{p^\mu}{M_\L}\,\g_5 + \a_4\,\g^\mu\g_5 
\end{bmatrix}u_\L(p,s).
\label{eq:vertexL}
\end{align}

The quark exchange kernel for the nucleon, $\Sigma$ and $\Xi$ reads
\begin{align}
Z_b &= 3
\begin{bmatrix} 
 \frac{1}{M_1}                  & \frac{1}{M_1}\g_\s\g_5         & -\frac{\sqrt{2}}{M_2}\g_\s\g_5 \\[0.8ex]
 \frac{1}{M_1}\g_5\g_\mu         & \frac{1}{M_1}\g_\s\g_\mu        & \frac{\sqrt{2}}{M_2}\g_\s\g_\mu \\[0.8ex]
-\frac{\sqrt{2}}{M_2}\g_5\g_\mu  & \frac{\sqrt{2}}{M_2}\g_\s\g_\mu & 0
\end{bmatrix}
\end{align}
where, in each case, $M_1$ is the mass of the singly represented dressed 
quark and $M_2$ the mass of the 
doubly represented dressed quark.\footnote{For the $\S^0$, the 
term ``doubly represented'' means the two light quarks of different flavours.} 
The factor of $3$ is obtained from projecting the kernel onto colour 
singlet states. For the $\L$ the quark exchange kernel is given by
%
%
%
\begin{align}
Z_\L &= 
\begin{bmatrix} 
0                                &  \frac{\sqrt{2}}{M_\ell}  & -\frac{\sqrt{2}}{M_\ell} \\[0.6ex] 
\frac{\sqrt{2}}{M_\ell}           & -\frac{1}{M_s}           & -\frac{1}{M_s}\g_\s\g_5 \\[0.6ex]
-\frac{\sqrt{2}}{M_\ell}\g_5\g_\mu &-\frac{1}{M_s}\g_5\g_\mu   & -\frac{1}{M_s}\g_\s\g_{\mu}
\end{bmatrix},
\end{align}
where $M_\ell$ is the mass of the dressed light quark. 
The quark--diquark bubble diagram matrix for the 
nucleon, $\Sigma$ and $\Xi$ reads
\begin{align}
\Pi_b(p) = 
\begin{bmatrix} 
\Pi_{q_2[q_1q_2]}(p)  & 0                       &  0  \\[0.8ex]
0               & \Pi_{q_2\{q_1q_2\}}^{\s\nu}(p) &  0  \\[0.8ex]
0               & 0                       & \Pi_{q_1\{q_2q_2\}}^{\s\nu}(p)
\end{bmatrix},
\end{align}
where for each baryon $q_1$ is the singly represented dressed quark 
and $q_2$ the doubly represented
dressed quark. For the $\L$ the analogous quantity reads
\begin{align}
\Pi_\L(p) = 
\begin{bmatrix} 
\Pi_{s[\ell\ell]}(p)  & 0               &  0  \\[0.8ex]
0                 & \Pi_{\ell[\ell s]}(p) &  0  \\[0.8ex]
0                 & 0               & \Pi_{\ell\{\ell s\}}^{\s\nu}(p)
\end{bmatrix}.
\end{align}
The quark--diquark bubble diagrams are given by
\begin{align}
\Pi_{q_i[q_jq_k]}(p) &= \int\frac{d^{4}k}{\left(2\pi\right)^{4}}\ S_{q_i}(k)\,\tau_{[q_j q_k]}(p-k), \\
\Pi_{q_i\{q_jq_k\}}^{\mu\nu}(p) &= \int\frac{d^{4}k}{\left(2\pi\right)^{4}}\ S_{q_i}(k)\,\tau^{\mu\nu}_{\{q_j q_k\}}(p-k).
\end{align}
Finally, the vertex normalization is given by
\begin{align}
\mathcal{Z}_B^{-1} = \lf.\overline{\G}_{0B}\ \frac{\partial\Pi_B(p)}{\partial p^2}\ \G_{0B}\rg|_{p^2 = M_B^2}.
\end{align}
Note, the value of $p^2$ which satisfies the Faddeev equation for each 
octet baryon defines its mass, $M_B^2$, 
and at that point the coefficients $\a_i$ then define the 
octet baryon vertex function.

\section{Results For Octet Baryons Masses\label{Parameters}}
%
\begin{table}[tbp]
\addtolength{\tabcolsep}{5.0pt}
\addtolength{\extrarowheight}{2.2pt}
\begin{tabular}{cccccccc}
\hline\hline
$\L_{IR}$ & $\L_{UV}$ & $M_\ell$  & $M_s$  & $G_\pi$ & $G_\rho$ & $G_s$  & $G_a$ \\
\hline
0.240    & 0.645    & 0.40   & 0.59   & 19.0   & 10.8  & 7.6    & 2.6    \\
\hline\hline
\end{tabular}
\caption{Model parameters are constrained to reproduce the physical pion 
and $\rho$ meson masses; the pion decay constant; the nucleon and $\Xi$ 
masses and the nucleon axial coupling. The infrared regulator and the 
dressed $u$ and $d$ quark masses -- labelled by $M_\ell$ -- are assigned 
their values \textit{a priori}. The regularization parameters and 
dressed quark mass are in units of GeV, while the couplings are in 
units of GeV\,$^{-2}$.} 
\label{tab:parameters}
\end{table}
%
\begin{table}[bp]
\addtolength{\tabcolsep}{8.0pt}
\addtolength{\extrarowheight}{2.2pt}
\begin{tabular}{ccccccc}
\hline\hline
$M_K$ & $M_{[\ell\ell]}$ & $M_{[\ell s]}$ & $M_{\{\ell\ell\}}$ & $M_{\{\ell s\}}$ & $M_{\{ss\}}$ \\[0.2em]
\hline
0.47  & 0.68          & 0.85         & 1.04           & 1.17           & 1.30       \\
\hline\hline
\end{tabular}
\caption{Results for the kaon mass, together with the various diquark masses, 
where the subscript $\ell = u,\,d$. Recall that the square brackets label 
scalar diquarks and the curly brackets axialvector diquarks. 
All masses are in units of GeV.} 
\label{tab:masses}
\end{table}
%
The NJL model employed in this work has the following parameters: 
two regularization parameters $\L_{IR}$ and $\L_{UV}$; 
the dressed quark masses $M_u$, $M_d$ and $M_s$;
\footnote{Alternatively, one may use the current quark masses 
$m_u$, $m_d$, $m_s$ from the NJL Lagrangian, as the gap equation 
provides a one-to-one correspondence with the dressed masses.} 
two coupling constants, $G_\pi$ and $G_\rho$, from the $\bar{q}q$ NJL 
Lagrangian, given in Eq.~\eqref{eq:lagrangian}; and the two coupling 
constants $G_s$ and $G_a$ from the $qq$ NJL Lagrangian. 
The infrared cutoff implements confinement and therefore should be of 
the order of $\L_{\text{QCD}}$ and we choose $\L_{IR} = 0.240\,$GeV; for the light-quark dressed masses we choose $M_u = M_d = 0.4\,$GeV; the ultraviolet cutoff $\L_{UV}$ and the couplings $G_\pi$ and $G_\rho$ are constrained by the empirical values for the pion decay constant, the pion mass and the $\rho$ mass; the $qq$ couplings are chosen to reproduce the physical nucleon mass and the nucleon axial coupling constant. Finally, the dressed $s$-quark mass is fixed to reproduce the empirical mass of the  cascade baryon ($\Xi$). The resulting parameters are summarized in Tab.~\ref{tab:parameters}.

Results for the kaon mass together with the various diquark masses are 
given in Tab.~\ref{tab:masses}. The splitting between the various scalar 
diquarks and between the axialvector diquarks is the result of 
explicit $SU(3)_F$ breaking effects from the strange quark. 
The empirical light to strange current quark mass ratio in 
the $\overline{\text{MS}}$ regularization scheme is 
$m_s/m_q = 27.5 \pm 1.0$~\cite{Beringer:1900zz}, while we find  
$m_s/m_q = 21.7$. For the analogous dressed quark mass ratio we obtain $M_s/M_q \simeq 1.5$, which illustrates that effects from DCSB are very much suppressed for the heavier strange quark. For completeness we give in Tab.~\ref{tab:residues} the pole residues for the meson and diquark $t$-matrices.

\begin{table}[tbp]
\addtolength{\tabcolsep}{5.0pt}
\addtolength{\extrarowheight}{2.2pt}
\begin{tabular}{cccccccc}
\hline\hline
$Z_\pi$ & $Z_K$ & $Z_{[\ell\ell]}$ & $Z_{[\ell s]}$ & $Z_{\{\ell\ell\}}$ & $Z_{\{\ell s\}}$ & $Z_{\{ss\}}$ \\[0.2em]
\hline
17.8   & 29.6  & 14.8          & 16.4         & 3.56           & 3.93          & 4.29       \\
\hline\hline
\end{tabular}
\caption{Results for the pole residues in the various meson and diquark 
$t$-matrices (c.f. Eqs.~(\ref{eq:scalarpropagatorpoleform}) and 
(\ref{eq:axialpropagatorpoleform}) ).} 
\label{tab:residues}
\end{table}

\begin{table}[bp]
\addtolength{\tabcolsep}{9.5pt}
\addtolength{\extrarowheight}{2.2pt}
\begin{tabular}{cccccccc}
\hline\hline
      & $M_N$ & $M_\L$  & $M_\Sigma$ & $M_\Xi$    \\
\hline
NJL   & 0.940 & 1.120  & 1.234    & 1.318    \\
Experiment & 0.940 & 1.116  & 1.193    & 1.318    \\
\hline\hline
\end{tabular}
\caption{Results for octet baryon masses and the average experimental mass for the corresponding multiplet. All experimental masses have an error of at most $0.015$\% but usually it is much less~\cite{Beringer:1900zz}. Because we have $M_u = M_d$ the masses of each member of the various isospin multiplets are degenerate. Recall that the nucleon and $\Xi$ masses were used to determine two of our NJL model parameters. All masses are in GeV.} 
\label{tab:octetmasses}
\end{table}
The octet baryon masses obtained by solving the appropriate Faddeev 
equation, as discussed in Sect.~\ref{sec:baryons}, are given in 
Tab.~\ref{tab:octetmasses}. In the $SU(3)_F$ limit all octet baryon masses 
are degenerate and hence the mass splitting between octet baryons results 
solely from the heavier $s$ quark mass. The mass splitting between the 
$\L$ and $\Sigma$ baryons is a consequence of the different diquark 
correlations which dominate their wavefunctions. The $\L$ baryon contains 
two types of scalar diquark and one type of axialvector 
diquark -- $[\ell\ell]$, $[\ell s]$ and $\{\ell s\}$ -- while 
the $\Sigma$ baryon contains one type of scalar diquark and two types of 
axialvector diquark -- $[\ell s]$, $\{\ell s\}$ and $\{\ell \ell\}$. 
Scalar diquarks are more bound than axialvector diquarks -- 
because of their strong connection with the pion and DCSB -- 
and consequently we find that the $\L$ is approximately $110\,$MeV 
lighter than the $\Sigma$ baryon. This is in reasonable agreement with 
the empirical mass splitting of approximately $80\,$MeV. 
A more precise fit would also need to include chiral corrections.
The parameters defining the Faddeev vertex function for each member of 
the baryon octet are summarized in Tab.~\ref{tab:vertex}.

\begin{table}[tbp]
\addtolength{\tabcolsep}{6.0pt}
\addtolength{\extrarowheight}{2.2pt}
\begin{tabular}{c|ccccccc}
\hline\hline
        & $\a_1$ & $\a_2$ & $\a_3$ & $\a_4$ & $\a_5$    \\
\hline
nucleon & 0.418 & 0.013 &-0.259 &-0.018 & 0.366 \\
$\L$    & 0.364 & 0.278 &-0.016 & 0.440 & --    \\
$\S$    & 0.351 & 0.032 &-0.215 &-0.021 & 0.406 \\
$\Xi$   & 0.388 & 0.017 &-0.273 &-0.015 & 0.364 \\
\hline\hline
\end{tabular}
\caption{Numerical coefficients that define the Faddeev vertex functions for 
each member of the
baryon octet. The nucleon, $\S$ and $\Xi$ vertex functions have 
the form given in Eq.~\eqref{eq:vertex} and
the $\L$ vertex function is given in Eq.~\eqref{eq:vertexL}.} 
\label{tab:vertex}
\end{table}

\section{Axial charges\label{AxialCh}}
The axial charges of the baryons are important because they connect 
the strong and weak interactions. They are also related to the quark 
spin content of the baryons~\cite{Jaffe:1989jz}. In fact, assuming 
$SU(3)$-flavour symmetry, all octet baryon decays can be parametrized 
by just three quantities: the Cabbibo angle, $\theta_{C}$, and 
the $F$ and $D$ couplings~\cite{Gaillard:1984ny,Hogaasen:1995nn}. 

The axial current of an octet baryon has the form
\begin{align}
J^{\mu,a}_{5,\l'\l}(p',p) &= \lf<p',\l'\lf|\bar{\psi}_q\,\g^\mu\g_5\,\l_a\,\psi_q\rg|p,\l\rg>, \no \\
&= \bar{u}(p',\l')\Bigl[\g^\mu\g_5\,G_A(Q^2) \no \\
&\hs{-16mm}
+ \frac{q^\mu\g_5}{2\,M_B}\,G_P(Q^2)
+ \frac{i\s^{\mu\nu}q_\nu\,\g_5}{2\,M_B}\, G_T(Q^2)\Bigr]\l_a\,u(p,\l).
\end{align}
where $q = p' - p$ is the 4-momentum transfer, $Q^2 \equiv -q^2$ and $\lambda$, $\lambda'$ represent
the initial and final nucleon helicity, respectively. 
The scalar functions $G_A(Q^2)$, $G_P(Q^2)$ 
and $G_T(Q^2)$ label the axial, induced pseudoscalar and induced 
pseudotensor form factors, respectively.
In this work we restrict ourselves to the $q \to 0$ limit, 
where the current becomes
\begin{align}
J^{\mu,a}_{5,\l}(p,p) &= G_A(0)\,\bar{u}(p,\l)\,\g^\mu\g_5\,\l_a\,u(p,\l).
\end{align}

The flavour-triplet axial charge of an octet baryon, $g^B_A$, is given by the matrix element
\begin{align}
g^B_A \, s^\mu = \lf<B \lf|\bar{\psi}\,\g^\mu\g_5\,\l_3\,\psi\rg|B\rg> = 
\lf(\D u_B - \D d_B\rg) s^\mu \, , 
\end{align}
where $s^\mu = \bar{u}(p,\l)\,\g^\mu\g_5\,u(p,\l)$ is the spin-vector of the octet baryon. The quark-spin fractions of the baryon are defined by
\begin{align}
\D q_B \, s^\mu &= \bigl<B \bigl|\bar{\psi}\,\g^\mu\g_5\,\hat{P}_q\,\psi\bigr|B\bigr>,
\end{align}
where the $u$ and $d$ quark projection operators are given by
\begin{align}
\hat{P}_q &= \frac{1}{2}\lf(\frac{2}{3}\,\ident \pm \l_3 + 
\frac{1}{\sqrt{3}}\,\l_8\rg),
\end{align}
and the plus sign corresponds to the $u$ quark.
\begin{figure}[tbp]
\centering\includegraphics[width=\columnwidth,clip=true,angle=0]{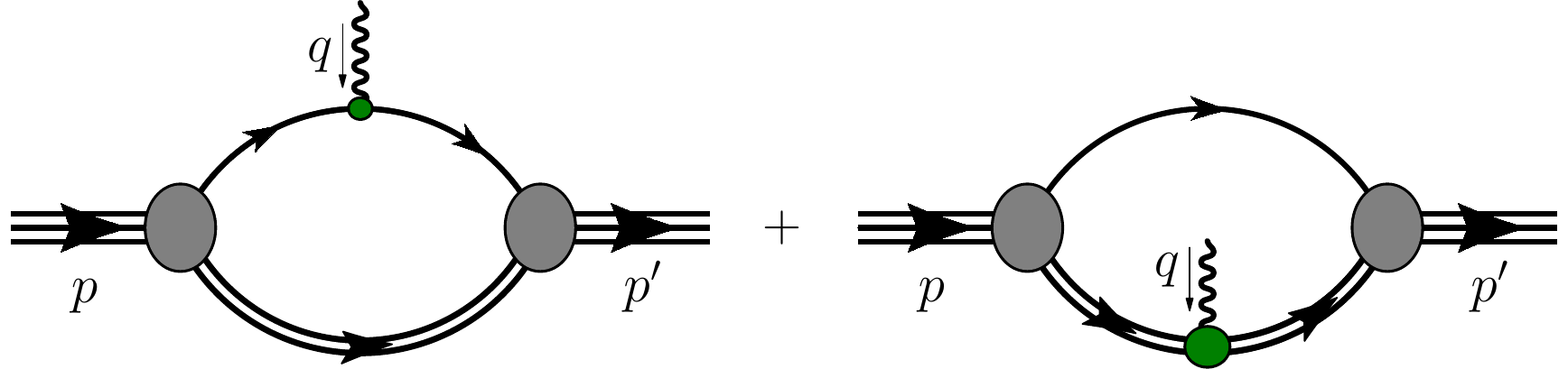}
\caption{(Colour online) Feynman diagrams representing the axial current for the octet baryons. The diagram on the left is called the quark diagram and the one on the right the diquark diagram. In the diquark diagram the photon interacts with each quark inside the non-pointlike diquark.}
\label{fig:nucleon_current}
\end{figure}

The various spin-fractions for the octet baryons under consideration are given by the sum of the six Feynman diagrams represented in Fig.~\ref{fig:nucleon_current} and have the structure
\begin{align}
&\D u_p = f^Q_{u[ud]} + f^Q_{u\{ud\}} \no \\
&\hs{3mm}
+ 2\,f^D_{u[ud]} + 2\,f^D_{u\{ud\}} + 4\,f^D_{d\{uu\}} + 2\,f^D_{u\{ud\}\leftrightarrow u[ud]}, \\
&\D d_p = f^Q_{d\{uu\}} + 2\,f^D_{u[ud]} + 2\,f^D_{u\{ud\}} - 2\,f^D_{u\{ud\}\leftrightarrow u[ud]}, \\[0.5ex]
&\D u_{\Sigma^{-}} = 0, \\
&\D d_{\Sigma^{-}} = f^Q_{d[ds]} + f^Q_{d\{ds\}} \no \\
&\hs{6mm}
+ 2\,f^D_{d[ds]} + 2f^D_{d\{ds\}} + 4\,f^D_{s\{dd\}} + 2\,f^D_{d\{ds\}\leftrightarrow d[ds]},  \\
&\D u_{\Xi^{-}} = 0, \\
&\Delta d_{\Xi^{-}} = f^Q_{d\{ss\}}
 + 2\,f^D_{s[ds]} + 2\,f^D_{s\{ds\}} - 2\,f^D_{s\{ds\}\leftrightarrow s[ds]}.  
\end{align}
The nomenclature for these Feynman diagrams is: a superscript $Q$ implies that the operator acts directly on a quark (\textit{quark diagram}) and a superscript $D$ implies that the operator acts on (a quark inside) a diquark (\textit{diquark diagram}); the notation $q_1[q_2 q_3]$ indicates a diagram with quark content $q_1 q_2 q_3$ contains only a scalar diquark of quark content $q_2 q_3$. Similarly the notation $q_1\{q_2 q_3\}$ indicates a diagram contains only an axialvector diquark of quark content $q_2 q_3$; and finally the notation $q_1\{q_2 q_3\} \leftrightarrow q_1[q_2 q_3]$
indicates the sum of the two diagrams where the operator induces a transition between scalar and axialvector diquarks
of flavour $q_2 q_3$. The numerical coefficients arise from the flavour structure of the operator and the Faddeev amplitude.

\begin{figure}[tbp]
\centering\includegraphics[width=\columnwidth,clip=true,angle=0]{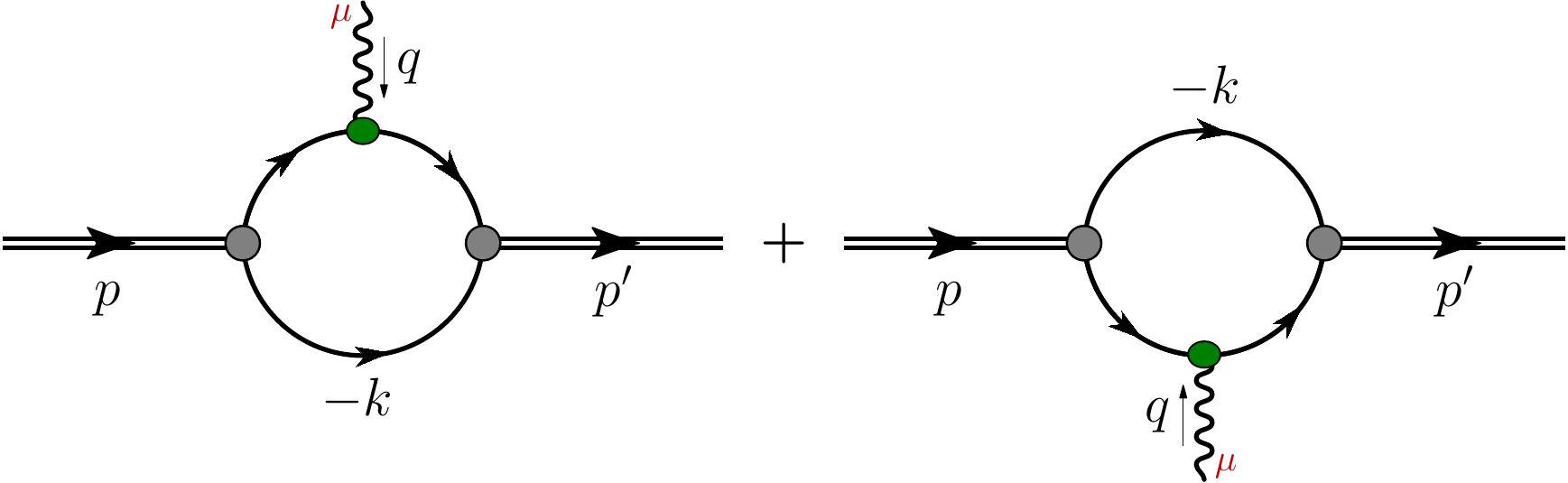}
\caption{(Colour online) Feynman diagrams that represent the diquark axial current. The shaded circles are the diquark Bethe-Salpeter vertices and the shaded oval is the $\g^\mu\g_5$ vertex.}
\label{fig:diquarkaxialcurrent}
\end{figure}

In general the quark diagram with a scalar diquark spectator reads
\begin{align}
&f^Q_{q_1[q_2 q_3]}\ \ol{u}\,\g^\mu\g_5\,u = 
\ol{\G}_{q_1[q_2 q_3]}(p) \int \frac{d^4k}{(2\pi)^4} \no \\
&\hs{5mm}
\times  S_{q_1}(k)\,\g^\mu\g_5\,S_{q_1}(k)\,\tau_{[q_2 q_3]}(p+k)\,\G_{q_1[q_2 q_3]}(p),
\end{align}
and the analogous diagram with an axialvector diquark spectator is given by
\begin{align}
&f^Q_{q_1\{q_2 q_3\}}\ \ol{u}\,\g^\mu\g_5\,u = 
\ol{\G}_{q_1\{q_2 q_3\},\a}(p) \int \frac{d^4k}{(2\pi)^4} \no \\
&\hs{1mm}
\times  S_{q_1}(k)\,\g^\mu\g_5\,S_{q_1}(k)\,\tau_{\{q_2 q_3\}}^{\a\b}(p+k)\,\G_{q_1\{q_2 q_3\},\b}(p).
\end{align}
Similarly, the general form of the diquark diagram with a scalar diquark reads
\begin{align}
&f^D_{q_1[q_2 q_3]}\ \ol{u}\,\g^\mu\g_5\,u = 
\ol{\G}_{q_1[q_2 q_3]}(p) \int \frac{d^4k}{(2\pi)^4}\ iS_{q_1}(p+k) \no \\
&\hs{13mm}
\times \tau_{[q_2 q_3]}(k)\ \L^\mu_{[q_2 q_3]}\ \tau_{[q_2 q_3]}(k)\,\G_{q_1[q_2 q_3]}(p),
\end{align}
and the analogous diagram with an axialvector diquark is given by
\begin{align}
&f^D_{q_1\{q_2 q_3\}}\ \ol{u}\,\g^\mu\g_5\,u = 
\ol{\G}_{q_1\{q_2 q_3\},\a}(p) \int \frac{d^4k}{(2\pi)^4}\ iS_{q_1}(p+k) \no \\
&\hs{4mm}
\times \tau^{\a\s}_{\{q_2 q_3\}}(k)\ \L^\mu_{\s\eta,\{q_2 q_3\}}\ \tau^{\eta\b}_{\{q_2 q_3\}}(k)\,\G_{q_1\{q_2 q_3\},\nu}(p),
\end{align}
where $\L^\mu_{[q_2 q_3]}$ and $\L^\mu_{\a\b,\{q_2 q_3\}}$ represent, respectively, the interaction of a scalar and 
axialvector diquark, with an axialvector current, in the $q \to 0$ 
and on-shell limits. Because the scalar diquark has spin-zero, 
we have $\L^\mu_{[q_2 q_3]} = 0$, while for the axialvector diquark we have
\begin{align}
\label{eq:axialvectordiquark}
\L^{\mu,\a\b}_{\{q_2 q_3\}}(p) &= 3i \int \frac{d^4k}{(2\pi)^4} \no \\
&\hs{-6mm}
\times \mathrm{Tr}_D\lf[\g^\b\,S(p+k)\,\g^\mu\g_5\,S(p+k)\,\g^\a\,S(k)\rg], \\
&= g_A^{\{q_2 q_3\}}\ i\ve^{\mu\a\b\l}p_\l \, ,
\end{align}
where $\a$ is the inital and $\b$ the final diquark polarization. 
The Feynman diagram for this contribution is illustrated 
in Fig.~\ref{fig:diquarkaxialcurrent}.
For the various axialvector diquarks we find:
$g_A^{\{\ell \ell\}} = -0.216$, $g_A^{\{\underline{\ell} s\}} = -0.194$, 
$g_A^{\{\ell \underline{s}\}} = -0.213$ and $g_A^{\{s s\}} = -0.194$\footnote{In our notation the underlined character denotes the quark that interacts with the photon. For example in $\{\underline{\ell} s\}$ the light quark interacts with the photon in the axialvector diquark.}. Note that in evaluating Eq.~\eqref{eq:axialvectordiquark}
we have used the on-shell condition, $\ve_\a(p)\,p^\a = 0$, for both the inital and final axialvector diquark.

The final Feynman diagram represents the mixing between the scalar and axialvector diquarks induced by the axial current, this diagram reads
\begin{align}
f^D_{q_1[q_2 q_3] \leftrightarrow q_1\{q_2 q_3\}} &\equiv f^D_{q_1[q_2 q_3] \to q_1\{q_2 q_3\}} + f^D_{q_1\{q_2 q_3\} \to q_1[q_2 q_3]},
\end{align}
where each contribution is given by\footnote{Note, when calculating these diagrams we in practice consider a small momentum transfer so that we can correctly identify $p'^2$ and $p^2$ with the inital and final diquark mass squared.}
\begin{align}
&f^D_{q_1[q_2 q_3] \to q_1\{q_2 q_3\}}\ \ol{u}\,\g^\mu\g_5\,u = 
\ol{\G}_{q_1\{q_2 q_3\},\a}(p)\ i\int \frac{d^4k}{(2\pi)^4} \no \\
&\hs{0mm}
S_{q_1}(p+k)\,\tau^{\a\s}_{\{q_2 q_3\}}(k)\ \L^\mu_{\s,[q_2 q_3]\to\{q_2 q_3\}}\ \tau_{[q_2 q_3]}(k)\,\G_{q_1[q_2 q_3]}(p),\\
&f^D_{q_1\{q_2 q_3\} \to q_1[q_2 q_3]} \ \ol{u}\,\g^\mu\g_5\,u = 
\ol{\G}_{q_1[q_2 q_3]}(p) \ i\int \frac{d^4k}{(2\pi)^4}\ \no \\
&\hs{0mm}
S_{q_1}(p+k)\, \tau_{[q_2 q_3]}(k)\ \L^\mu_{\s,\{q_2 q_3\}\to[q_2 q_3]}\ \tau^{\s\a}_{\{q_2 q_3\}}(k)\,\G_{q_1\{q_2 q_3\},\a}(p).
\end{align}
The diquark transition vertices are given by
\begin{align}
\L^{\mu\a}_{[q_2 q_3]\to\{q_2 q_3\}} &=  3i \int \frac{d^4k}{(2\pi)^4} \no \\
&\hs{-10mm}
\times \mathrm{Tr}_D\lf[\g_5\,S(p+k)\,\g^\mu\g_5\,S(p+k)\,\g^\a\,S(k)\rg],\\
\L^{\mu\a}_{\{q_2 q_3\}\to[q_2 q_3]} &= 3i \int \frac{d^4k}{(2\pi)^4} \no \\
&\hs{-10mm}
\times \mathrm{Tr}_D\lf[\g^\a\,S(p+k)\,\g^\mu\g_5\,S(p+k)\,\g_5\,S(k)\rg].
\end{align}
These vertices have the general form
\begin{align}
\L^{\mu\a}_{[q_2 q_3]\to\{q_2 q_3\}} &= a_{q_2 q_3}\, g^{\mu\a} + b_{q_2 q_3}\, p^\mu p^\a
                             = - \L^{\mu\a}_{\{q_2 q_3\}\to[q_2 q_3]}.
\end{align}
For the various diquark transitions we find: $a_{\ell\ell} = -0.054$, $a_{\underline{\ell} s} = -0.052$, $a_{\ell \underline{s}} = -0.048$, $b_{\ell\ell} = 0.092$, $b_{\underline{\ell} s} = 0.096$ and $b_{\ell \underline{s}} = 0.115$. 

\begin{table}[tbp]
\addtolength{\tabcolsep}{6.9pt}
\addtolength{\extrarowheight}{2.2pt}
\begin{tabular}{cccccccc}
\hline\hline
$\D u_n$ & $\D d_n$ & $\D u_{\Sigma^-}$ & $\D d_{\Sigma^-}$ & $\D u_{\Xi^-}$ & $\D d_{\Xi^-}$ \\[0.2em]
\hline
1.145  &  0.331  &  0  &   -1.048  &  0  &  0.313  \\
\hline\hline
\end{tabular}
\caption{Results for the spin fractions in the nucleon, $\Sigma$, and $\Xi$.} 
\label{tab:spinfractions}
\end{table}

\begin{figure}[tbp]
\centering\includegraphics[width=\columnwidth,clip=true,angle=0]{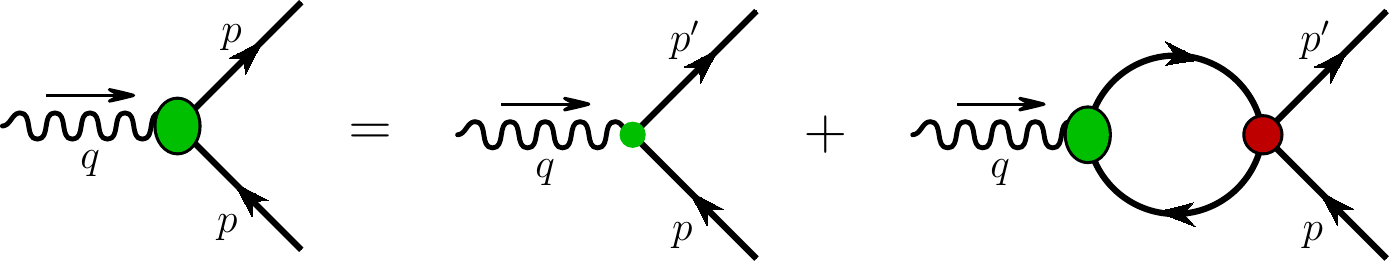}
\caption{(Colour online) Inhomogeneous Bethe-Salpeter equation whose solution gives the quark-axialvector vertex, represented as the large shaded oval. The small dot is the inhomogeneous driving term, while the shaded circle is the $\bar{q}q$ interaction kernel. Only the $\rho$ interaction channel contributes.}
\label{fig:quarkaxialvertex}
\end{figure}

Evaluating these diagrams gives the spin-fractions which we summarize in Tab.~\ref{tab:spinfractions}.
In addition to these body form factor contributions, the axial charge of the quark receives a finite renormalization. 
This renormalization is given by the inhomogeneous BSE illustrated in Fig.~\ref{fig:quarkaxialvertex}. The renormalized axial charge of the light quark is given by
\begin{align}
g_A^q = \frac{1}{1 + 2\,G_{a_1}\,\Pi_{AA}^{(T)}(0)},
\end{align}
where $\Pi_{AA}^{(T)}(q^2)$ is the transverse piece of the bubble diagram
\begin{align}
\Pi^{\mu\nu}_{AA}(q^2) &= 6i\int \frac{d^4k}{(2\pi)^4} 
\mathrm{Tr}_D\lf[\g^\mu\g_5\,S(k+q)\,\g^\nu\g_5\,S(k)\rg], \no \\
&\hs{-6mm}
\equiv \lf(g^{\mu\nu} - \frac{q^\mu q^\nu}{q^2}\rg)\Pi_{AA}^{(T)}(q^2) + \frac{q^\mu q^\nu}{q^2}\ \Pi_{AA}^{(L)}(q^2).
\end{align}
The coupling $G_{a_1}$ is adjusted $(G_{a_1} = 1.0)$ to give $M_{a_1} = 1.26$ GeV. The unrenormalized quark axial charge is unity, however for the renormalized axial charge we find $g_A^q = 0.935$.
The value of the axial charge for octet baryon strangeness conserving transitions, for the bare case (``Bare'') and for the case with a renormalized axial quark vertex (``BSE'') are given in Tab.~\ref{tab:results}.

The axial charges for the octet baryons also receive corrections from the 
meson cloud. In order to preserve the correct non-analytic behavior, required
by chiral symmetry, it is necessary to compute these at the hadronic 
level~\cite{Thomas:1999mu,Thomas:2000fa}. This requires the meson-baryon
coupling constants and form factors. The former are given by the appropiate
Goldberger-Treiman relation, which is respected because the NJL model respects
chiral symmetry. The latter are related to the hadron sizes. Since both the
axial charges and hadron sizes in the present model are very close to those
calculated in the cloudy bag model (CBM)~\cite{Theberge:1980ye,Thomas:1982kv},
we can take the meson corrections directly from the work of Kubodera et 
al.~\cite{Kubodera:1984qd}. In practice this means 
that our results should be multiplied by 0.92 for $g^n_A$, 0.90 for 
$g^{\Sigma}_A$ and 0.95 for $g^{\Xi}_A$. 
Our final results, including corrections from the quark vertex 
renormalization and the meson cloud are shown in Tab.~\ref{tab:results}. 
The only experimental value we have available is $g_{A}^{n}$. 
The values of the vector form factors $f_{1}(0)$ are 1, $\sqrt{2}$ 
and 1 for the nucleon, $\Sigma$, and $\Xi$, respectively. This comes from 
charge conservation when the electromagnetic form factors are computed 
in the octet. 

We are now in a position to determine the size of SU(3)-flavour breaking 
effects for the axial charges of the octet baryons, using 
the SU(3)-flavour parametrization of Ref.~\cite{Gaillard:1984ny}, namely
\begin{align}
n &\to p + \nu_e + e^-  &\Longrightarrow && \left(g_A/f_1\right)^n &= F+D, \\
\Sigma^- &\to \Sigma^0 + \nu_e + e^-  &\Longrightarrow && \left(g_A/f_1\right)^{\Sigma}&= F, \\
\Xi^- & \to \Xi^{0} + \nu_e + e^- &\Longrightarrow && \left(g_A/f_1\right)^{\Xi} &= F-D.
\end{align}
%
\begin{table}[tbp]
\addtolength{\tabcolsep}{2.5pt}
\addtolength{\extrarowheight}{2.2pt}
\begin{tabular}{cccccccc}
\hline\hline
            & Bare  & BSE  & BSE + CBM & exp't/Ref.~\cite{Yamaguchi:1989sx}          \\[0.2em]
\hline
$\left(g_A/f_1\right)^{n}$      & 1.48    & 1.38   & 1.27       & 1.2701 $\pm$ 0.0025   \\
$\left(g_A/f_1\right)^{\Sigma}$  & 0.52    & 0.49   & 0.44       &  0.44    \\
$\left(g_A/f_1\right)^{\Xi}$    & - 0.31  & - 0.29 & - 0.28     &  - 0.32    \\
\hline\hline
\end{tabular}
\caption{Axial charges for the different beta decays with $\Delta S=0$ 
in three cases: Bare, treats the quark as a point-like particle, 
without structure and without a meson cloud; BSE, includes the 
renormalization of the quark axial charge through the solution of the BSE 
and, finally, including the BSE renormalization as well as meson cloud 
corrections computed within the CBM. As the experimental result is only 
known for the nucleon case, for the $\Sigma$ and $\Xi$ we show the 
results from the CBM computation in Ref.~\cite{Yamaguchi:1989sx}, modified 
very slightly to reproduce the current experimental value of $g_A^n$.} 
\label{tab:results}
\end{table}
%
Within our model the values of $F$ and $D$ may be computed by choosing any 
pair of the previous relations. We call $F_{\Sigma(\Xi)}$ 
and $D_{\Sigma(\Xi)}$ the parameters calculated from 
$\left(g_A/f_1\right)^{\Sigma(\Xi)}$ and $\left(g_A/f_1\right)^n$. 
{}From Tab.~\ref{tab:results} we obtain
\begin{align}
F_{\Sigma} &= \left(g_A/f_1\right)^{\Sigma} = 0.441, \nonumber\\ 
D_{\Sigma} &= \left(g_A/f_1\right)^n-\left(g_A/f_1\right)^{\Sigma} = 0.829,
\end{align}
and
\begin{align}
F_{\Xi} &= \tfrac{1}{2}\boldsymbol{(}\left(g_A/f_1\right)^{n} + \left(g_A/f_1\right)^{\Xi}\boldsymbol{)} = 0.496, \nonumber\\ 
D_{\Xi} &= \tfrac{1}{2}\boldsymbol{(}\left(g_A/f_1\right)^{n} - 
\left(g_A/f_1\right)^{\Xi}\boldsymbol{)} = 0.774 \, .
\end{align}
The discrepancies suggest SU(3)-flavour symmetry breaking effects of 
around 10\%, with $F_{\Sigma}/F_{\Xi} = 0.89 $ and 
$D_{\Sigma}/D_{\Xi} = 1.07$. Since the strangeness conserving 
$\beta$-decays for the $\Sigma^{-}$ and $\Xi^{-}$ have not yet been measured, 
this result should be viewed as a prediction to be tested experimentally.
We note that even larger SU(3) violation has been reported in the 
context of the proton spin problem~\cite{Bass:2009ed}.

In addition, a comparison of our results with the cloudy bag model 
computations performed in Ref.~\cite{Yamaguchi:1989sx} shows that 
$g_{A}^{\Sigma}$ is the same in both models, whereas $g_{A}^{\Xi}$ is 
slightly smaller in magnitude in our work. 
The calculation of Ref.~\cite{Yamaguchi:1989sx} includes 
one-gluon exchange and center of mass corrections, plus recoil effects 
and a rescaling factor to reproduce the experimental $g_{A}^{n}$.

The other source of ``data'' with which our model might be compared 
are lattice QCD calculations.
There has recently been good progress in the calculation of the 
electromagnetic form factors for the octet 
baryons~\cite{Shanahan:2014cga,Shanahan:2014uka}, using chiral 
extrapolations of the lattice results.
Clearly an extension of that work to weak form factors would 
provide a valuable test our model predictions.

\section{CONCLUSIONS\label{sect:con}}

We have computed the masses and the $\Delta S = 0$ axial charges 
of the baryon octet using a confining NJL model. The model results for 
the masses are in good agreement with the experimental values. While there are 
currently no measurements of the $\Delta S = 0$ axial charges, 
other than for the 
neutron, we did find very close agreement between our results and those 
found within the cloudy bag model. 
Since there is currently considerable discussion 
concerning the degree of violation of 
SU(3) symmetry, the deviation of order 10\% which we found is 
significant.

It will be important to extend the present investigation 
to calculate the chiral corrections explicitly within this model.  
Given the new lattice results for octet baryon electromagnetic form factors,
we look forward to simulations of a similar quality for the axial form factors.
Meantime, it would be very interesting to extend the present model 
to calculate the hyperon electromagnetic form factors.

\section*{ACKNOWLEDGEMENTS}
ICC thanks Wolfgang Bentz invaluable conversations. 
This material is based upon work supported by the U.S. Department of Energy, Office of Science, Office of Nuclear Physics, under contract number DE-AC02-06CH11357; and the Australian Research Council through the 
ARC Centre of Excellence in Particle Physics at the Terascale 
and an ARC Australian Laureate Fellowship FL0992247 (AWT).


\end{document}